\documentclass[aps, pra, preprint, english]{revtex4-1}

\usepackage[utf8]{inputenc}

\usepackage{mathptmx}      
\usepackage{latexsym}

\usepackage[T1]{fontenc}

	\usepackage[intlimits]{amsmath}

	\usepackage{fixmath}

	\usepackage{amssymb}

	\usepackage{mathtools}
	\mathtoolsset{showonlyrefs}

	\usepackage[caption=false]{subfig}

	\usepackage{dsfont}

	\usepackage{graphicx}
	
	\usepackage[final,colorlinks=true,linkcolor=blue,citecolor=blue]{hyperref}

\renewcommand*\Re{\operatorname{Re}}
 
  \renewcommand{\Upsilon}{\varUpsilon}

  \newcommand*\setS{\mathcal{S}}
  \newcommand*\setD{\mathcal{D}}

  \newcommand*\Ndets{N}
  \newcommand*\permut{\sigma}
  
  \newcommand*\detd{d}

  \newcommand*\SymmGroup[1]{\Sigma_{#1}}
  
  \newcommand*\BSPindex{av}

  \newcommand*\TotalProb[1]{P_{\text{av}}(\setD;\setS)}
  \newcommand*\ProdMatrix[2]{\mathcal{A}^{(\setD,\setS)}_{#1}}
  \newcommand*\OverlapFrequency[2]{g(#1,#2)}
  \newcommand*\OverlapFactor[1]{f_{#1}(\setS)}
  \newcommand*\TotalProbIntervalSingle[1]{P\left( \tdettintervalSet, \setD; \setS \right)}

  \newcommand*{\Umatrix}[1]{\mathcal{U}^{(\setD,\setS)}_{#1}}

  \newcommand*\Uto[2]{\mathcal{U}_{#2,#1}}
  \newcommand*\Utoconj[2]{\conj{\mathcal{U}}_{#2,#1}}

  \newcommand*\tdet[1]{t_{#1}}

  \newcommand*\tinterval[1]{\Delta \tdet{#1}}
  \newcommand*\tdettintervalSet{ \left\{ \tdet{\detd}, \tinterval{\detd} \right\}}

							\let\DeltaDefault\Delta
							\renewcommand*\Delta{\DeltaDefault\!}

  \newcommand*{\unity}{\mathds{1}}
  \newcommand{\conj}[1]{#1^*}
  \newcommand{\ii}{\mathrm{i}}
  \newcommand*{\per}{\operatorname{perm}}
  \newcommand{\abs}[1]{\left\lvert{#1}\right\rvert}
  \newcommand{\ee}[1]{\operatorname{e}^{#1}}
  \newcommand*{\defeq}{\mathrel{\vcenter{\baselineskip0.5ex \lineskiplimit0pt
                     \hbox{\scriptsize.}\hbox{\scriptsize.}}}%
                     =}
  \renewcommand{\Omega}{\varOmega}
  \renewcommand*\vec[1]{\mathbold{#1}}
  
  \newcommand{\ket}[1]{\vert #1 \rangle}

  \newcommand{\matrixel}[3]{\langle #1 \vert #2 \vert #3 \rangle}

  \renewcommand\d[1]{d #1\,}
\begin{document}

\title{Multi-Boson Correlation Sampling}
	\author{Vincenzo Tamma}
	\email{vincenzo.tamma@uni-ulm.de}
	\author{Simon Laibacher}
	\affiliation{Institut f\"{u}r Quantenphysik and Center for Integrated Quantum Science and Technology (IQ\textsuperscript{ST}), Universit\"{a}t Ulm, Albert-Einstein-Allee 11, D-89081 Ulm, Germany}

\begin{abstract}
We give a full description of the problem of \emph{Multi-Boson Correlation
Sampling} (MBCS) at the output of a random interferometer for single input
photons in arbitrary multimode pure states. The MBCS problem is the task of
sampling  at the interferometer output from the probability distribution
associated with polarization- and time-resolved detections. We discuss
the richness of the physics and the complexity of the MBCS problem for
nonidentical input photons. We also compare the MBCS problem with the standard boson sampling
problem, where the input photons are assumed to be identical and the system is
``classically'' averaging over the detection times and polarizations.
\end{abstract}
\maketitle

	\begin{center}
		In memory of Dr. Howard Brandt
	\end{center}
\section{Motivation}
Multi-Boson interference based on correlated measurements is at the
heart of many fundamental phenomena in quantum optics
and of numerous applications in quantum information \cite{Pan2012,TammaLaibacherPRL}.
Recent works
\cite{Yao2012,Ra2013,Metcalf2013,Broome2013,Crespi2013,tillmann2013experimental,Tillmann2014,Spring2013,Shen2014,Bentivegna2015} have demonstrated the
feasibility of multi-boson experiments based on higher-order correlation
measurements well beyond the first two-boson interference experiments \cite{Alley1986,Shih1988,Hong1987}. In particular, a lot of attention
in the research community was drawn to the so-called boson sampling problem \cite{aaronson2011computational,Franson2013,Ralph2013}, formulated as the task to sample from the
probability distribution of finding $N$ single input bosons at the output of a passive linear interferometer. This probability distribution depends on permanents of random complex matrices. The matrix permanent is defined similarly to the determinant aside from the minus signs appearing in the determinant. However, differently from the determinant, the permanent cannot be calculated \cite{Valiant1979} or even approximated \cite{aaronson2011computational} in polynomial time by a classical computer 
(more precisely these computational problems are in the complexity class \#P).
It has been argued that this also implies that the boson sampling problem cannot be solved efficiently by a classical computer \cite{aaronson2011computational}.

This result has triggered several multi-boson interference experiments
\cite{Broome2013,Crespi2013,tillmann2013experimental,Tillmann2014,Spring2013,Shen2014} as well as
studies of its characterization
\cite{Spagnolo2014,Carolan2013,Tichy2014,deGuise2014,Tan2013}.

In its current formulation, the boson sampling problem relies only on sampling over all possible
subsets of detected output ports regardless of the time and the polarization associated with each detection.
However,  thanks to the modern fast detectors and the possibility of producing single photons with arbitrary temporal and spectral properties \cite{Keller2004,Kolchin2008,Polycarpou2012} 
time-resolved correlation measurements \cite{TammaLaibacherPRL,Legero2004,Tamma2013,Zhao2014} are today at hand  experimentally. 

This has motivated us to introduce the novel problem of Multi-Boson Correlation Sampling (MBCS) \cite{TammaLaibacherPRL,LaibacherTammaComplexity}, which considers the sampling process from the interferometer output probability distribution depending on the 
output ports where the photons are detected and the corresponding detection times and polarizations.

The paper is organized as follows: In section \ref{sec:MBCSP} we give a general
description of the MBCS problem for correlation measurements of arbitrary order $N$ in a $2M \times 2M$-port interferometer including the case of photon bunching at the detectors. In section~\ref{sec:MBCSrates}, we then analyze the multi-photon indistinguishability at the detectors in the context of time- and polarization-resolved detections. In the limit $N\ll M$ where no bunching occurs, we discuss the degree of multi-photon correlation interference for different scenarios of multi-photon distinguishability in section~\ref{sec:MBCP11}.

In section~\ref{sec:Integrated}, we describe the case of detectors averaging over the detection times and polarizations. We consider again the limit  $N \ll M$ in section~\ref{sec:Integrated1} and show in section \ref{sec:BSP} that, for identical input photons, this scenario corresponds to the description of the well known boson sampling problem. 

\section{Multi-Boson Correlation Sampling (MBCS)}\label{sec:MBCSP}

\begin{figure} 
  \begin{center}
	  \subfloat[MBCS with bunching]{
	  \includegraphics[]{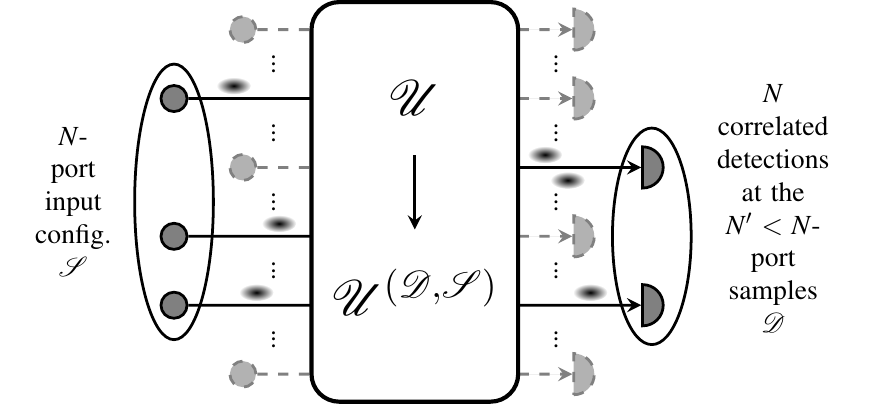}
  }

  \subfloat[MBCS without bunching]{
	  \includegraphics[]{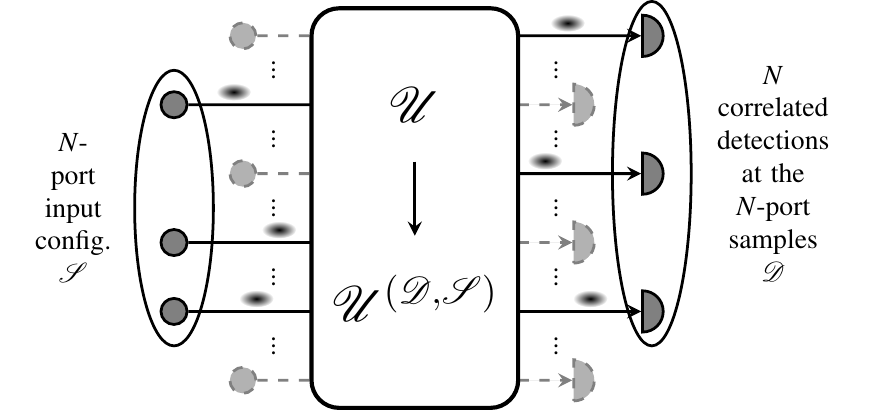} 
  }
  \caption{Multi-boson correlation sampling setup, given by a linear, passive interferometer with $2M \times 2M$ ports that is fully characterized by a unitary $M\times M$ matrix $\mathcal{U}$. The sampling is performed by injecting $N$ single bosons into $N$ of the $M$ input ports, the set of which we label by $\mathcal{S}$, and detecting
  them at the output ports by time- and polarization-resolved detectors. A detection sample $(\mathcal{D},\{t_j,\vec{p}_j\}_{j=1,\dots,N})$ is defined by the set $\mathcal{D}$ of output ports where a detection occurs and by the corresponding detection times and polarizations $\{t_j,\vec{p}_j\}_{j=1,\dots,N}$. For fixed $\mathcal{S}$ and $\mathcal{D}$, the evolution through the interferometer is fully characterized by a $N\times N$ submatrix $\mathcal{U}^{(\mathcal{D},\mathcal{S})}$ of $\mathcal{U}$.
  In panel (a) we consider the general case where photon bunching at the detectors may occur, resulting in possible repetitions of the entries in $\mathcal{D}$, while in panel (b) we address the limit $N \ll M$ where photon-bunching events are negligible and all elements of $\mathcal{D}$ are different.}
	  \label{fig:InterferometerSetup}
  \end{center}
\end{figure}

Let us describe the physical problem of Multi-Boson Correlation Sampling (MBCS) \cite{TammaLaibacherPRL,LaibacherTammaComplexity}.  First, a random linear interferometer with $2M\geq2N$ ports is implemented (see Fig.~\ref{fig:InterferometerSetup}(a)). This requires only a polynomial number (in $M$) of passive linear
optical elements \cite{Reck1994}. $N$ single photons are then prepared
at $N$ input ports of the interferometer, where 
$$\mathcal{S} = (s_1,\dots,s_i,\dots,s_N)$$
(with $s_i =1,\dots,M$) is the chosen set of occupied input ports. In the following, we will label the operators at the occupied input ports $s_i$ only with the index $i$ to simplify the notation. The $N$-photon input state 
\begin{align}
	\ket{\mathcal{S}} \defeq \bigotimes_{i=1}^N
	\ket{1[\vec{\xi}_{i}]}_{s_i}
	\bigotimes_{s \notin \mathcal{S}}
	\ket{0}_{s},
	\label{eqn:StateDefinition}
\end{align}
is then defined, in a given polarization basis $\{\vec{e}_1,\vec{e}_2\}$, by the $N$ single-photon multimode states
\begin{align}
	\ket{1[\vec{\xi}_{i}]}_{s_i} \defeq \sum_{\lambda=1,2} \int_{0}^{\infty} \d{\omega} \left( \vec{e}_{\lambda}\cdot \vec{\xi}_{i}(\omega) \right) \hat{a}^{\dagger}_{i,\lambda} (\omega) \ket{0}_{s_i},
	\label{eqn:SinglePhotonState}
	\noeqref{eqn:SinglePhotonState}
\end{align}
with the creation operator $\hat{a}^{\dagger}_{i,\lambda}(\omega)$ associated with the port $s_i$,
the frequency mode $\omega$, and the polarization $\lambda$ \cite{Loudon2000}.
The complex spectral distribution
\begin{align}
	\vec{\xi}_i(\omega) \defeq \vec{v}_i \; \xi_i(\omega - \omega_{0i}) \ee{\ii \omega t_{0i}}
	\label{eqn:ComplexSpectralDistribution}
\end{align}
is defined by the polarization $\vec{v}_i$, the spectral shape $\xi_i(\omega) \in \mathds{R}$ with normalization condition
\begin{align}
	\int_{0}^{\infty}\d{\omega} \abs{\xi_i(\omega)}^2 = 1,
\end{align}
the color or central frequency $\omega_{0i}$, and the time $t_{0i}$ of emission of the photon injected in the port $s_i \in \mathcal{S}$.
By using $M$ detectors at the $M$ output ports of the interferometer, we consider the sampling process from all possible correlated detections of the $N$ input photons depending on the detection times and polarizations. In particular, the $N$ input photons can be detected in the output ports $d_j= 1,2,...,M\ (j=1,\dots,N)$, defining the 
port sample 
\begin{align}
\mathcal{D} \defeq ( d_1, \dots , d_j, \dots , d_N ),
\label{eqn:PortSample}
\end{align}
where a port $d =1,2,...,M$ is contained $n_d$ times if $n_d$ photons are detected in that port. Thereby, each possible measurement outcome is defined by the port sample $\mathcal{D}$ together with the respective detection times $\{t_j\}_{j=1,2,...N}$ and polarizations $\{\vec{p}_j\}_{j=1,2,...N}$, with $\vec{p}_j = \vec{e}_{1}, \vec{e}_{2}$.

\section{MBCS Probability Rates}\label{sec:MBCSrates}

We will now determine the $N$-photon probability rates for each possible multi-boson correlation sample $(\mathcal{D},\{t_j,\vec{p}_j\}_{j=1,\dots,N})$ defined by the port sample $\mathcal{D}$ and the respective detection times $\{t_j\}_{j=1,2,...N}$ and polarizations $\{\vec{p}_j\}_{j=1,2,...N}$.

We consider input photons with frequency spectra satisfying the narrow bandwidth approximation and a polarization-independent interferometric evolution with approximately equal propagation time $\Delta t$ for each possible path. Moreover,
we define the $N \times N$ matrix
\begin{align}
	\mathcal{U}^{(\mathcal{D},\mathcal{S})} = [ \mathcal{U}^{(\mathcal{D},\mathcal{S})}_{j,i} ]_{\substack{j=1,\dots,N \\ i=1,\dots,N}}
	\defeq [ \mathcal{U}_{d_j,s_i} ]_{\substack{j=1,\dots,N \\ i=1,\dots,N}}
\end{align}
obtained from the $M\times M$ unitary matrix $\mathcal{U}$ describing
the interferometer evolution. Here, the elements $\mathcal{U}^{(\mathcal{D},\mathcal{S})}_{j,i}$ of different rows can be the same if multiple photons are detected at the same port. The field operators $\hat{\vec{E}}^{(+)}_{d_j}(t_j)$ at the detected ports $d_j\in \mathcal{D}$ can then be written in terms of the operators $\hat{\vec{E}}^{(+)}_{i} (t_j-\Delta t)$ at the input ports $s_i\in \mathcal{S}$ as
\begin{align}
	\hat{\vec{E}}^{(+)}_{d_j}(t_j) = \sum_{i=1}^N
	\mathcal{U}^{(\mathcal{D},\mathcal{S})}_{j,i} \hat{\vec{E}}^{(+)}_{i} (t_j-\Delta t).
	\label{eqn:ExpansionAout}
\end{align}

We calculate now the rate of an $N$-fold detection event for ideal photodetectors given by the $N$th-order correlation function \cite{Glauber2007}
\begin{align}
	G^{(\mathcal{D},\mathcal{S})}_{\{t_j,\vec{p}_j\}} = \matrixel{\mathcal{S}}{\prod_{j=1}^{N} \Big(\vec{p}^{*}_j \cdot \hat{\vec{E}}^{(-)}_{d_j}(t_j) \Big) \prod_{j=1}^{N} \Big( \vec{p}_j \cdot \hat{\vec{E}}^{(+)}_{d_j}(t_j) \Big)}{\mathcal{S}},
	\label{eqn:ProbabilityRate}
\end{align}
where $\vec{p}_j\cdot \hat{\vec{E}}^{(+)}_{d_j} (t_j)$ is the component of the electric field operator in Eq. \eqref{eqn:ExpansionAout} in the detected polarization $\vec{p}_j$. From an experimental point of view, we can assume that for any sample $(\mathcal{D}, \{t_j,\vec{p}_j\}_{j=1,\dots,N})$ the integration time $T_I$ of the detectors is short enough that the probability rate in Eq.~\eqref{eqn:ProbabilityRate} remains constant during $T_I$.

By using Eq.~\eqref{eqn:ExpansionAout} and defining the $\Ndets\times\Ndets$
operator matrices
\begin{align}
	\hat{\mathcal{M}}^{(\mathcal{D},\mathcal{S})}_{\{t_j,\vec{p}_j\}} \defeq
	\Big[ \mathcal{U}^{(\mathcal{D},\mathcal{S})}_{j,i} \big( \vec{p}_j \cdot \hat{\vec{E}}^{(+)}_{i}(t_j - \Delta t) \big)
	\Big]_{\substack{j=1,\dots,N\\i=1,\dots,N}},
	\label{eqn:OperatorMatrixDef}
\end{align}
Eq.~\eqref{eqn:ProbabilityRate} reduces to (see App.~\ref{app:OperatorPermanents}) 
\begin{align}
	G^{(\mathcal{D},\mathcal{S})}_{\{t_j,\vec{p}_j\}} &=
	\matrixel{\setS}{ \Big( \sum_{\sigma'\in
	\Sigma_N}\prod_{j=1}^{N}
	\big( \mathcal{U}^{(\mathcal{D},\mathcal{S})}_{j,\sigma'(j)}\big)^{*} \big( \vec{p}^{*}_j \cdot \hat{\vec{E}}^{(-)}_{\sigma'(j)}(t_{j}-\Delta t) \big) \Big) \\
	& \hspace{2cm}\times \Big( \sum_{\sigma\in \Sigma_N} \prod_{j=1}^N \mathcal{U}^{(\mathcal{D},\mathcal{S})}_{j,\sigma(j)} \big( \vec{p}_j \cdot \hat{\vec{E}}^{(+)}_{\sigma(j)}(t_{j}-\Delta t) \big)
	\Big)}{\setS} \nonumber \\
	&= \matrixel{\setS}{ \big( \per \hat{\mathcal{M}}^{(\mathcal{D},\mathcal{S})}_{\{t_j,\vec{p}_j\}} \big)^{\dagger} \big( \per \hat{\mathcal{M}}^{(\mathcal{D},\mathcal{S})}_{\{t_j,\vec{p}_j\}} \big)
}{\setS},
	\label{eqn:CorrelationOperatorMatrix}
\end{align}
where $\Sigma_N$ is the symmetric group of order $N$ and we used the definition of matrix permanents \cite{Minc1984} 
\begin{align}
\per A \defeq \sum_{\sigma \in \Sigma_N} \prod_{i=1}^N A_{i,\sigma(i)}.
\label{eqn:PermDef} 
\end{align}
In Eq. (\ref{eqn:CorrelationOperatorMatrix}) the permanent structure of the correlation function emerges already in terms of the operators contributing to the expectation value for each given sample $(\mathcal{D},\{t_j,\vec{p}_j\}_{j=1,\dots,N})$.

Further, by using the Fourier transforms 
\begin{align}
	\vec{\chi}_i(t)\defeq \mathcal{F}[\vec{\xi}_i](t-\Delta t) = \vec{v}_i \, \chi_i(t-t_{0i}-\Delta t) \ee{\ii \omega_{0i}(t-t_{0i}-\Delta t)}
	\label{eqn:Fourier}
\end{align}
of the frequency distributions $\vec{\xi}_i$, where $\chi_i(t)$ is the Fourier transform of $\xi_i(\omega)$ in Eq.~\eqref{eqn:ComplexSpectralDistribution}, we define the matrices
\begin{align}
	\mathcal{T}^{(\mathcal{D},\mathcal{S})}_{\{t_j,\vec{p}_j\}} \defeq [ \mathcal{U}^{(\mathcal{D},\mathcal{S})}_{j,i} \big(\vec{p}_j \cdot \vec{\chi}_i(t_j) \big) ]_{\begin{array}{l}\scriptstyle j=1,\dots,N \\[-0.2em]\scriptstyle i=1,\dots,N 
		\label{eqn:CorrMatrixDefintion}
		\end{array}}.
\end{align}
Here, for each multi-boson correlation sample $(\mathcal{D},\{t_j,\vec{p}_j\}_{j=1,\dots,N})$, each matrix entry $\mathcal{U}^{(\mathcal{D},\mathcal{S})}_{j,i} \big(\vec{p}_j \cdot \vec{\chi}_i(t_j) \big)$ describes the  probability amplitude for the quantum path from the source port $s_i$ to the port $d_j$ where a single-photon detection occurs at time $t_j$ with polarization $\vec{p}_j$. 
Each possible product $$ \prod_{j=1}^N \mathcal{U}^{(\mathcal{D},\mathcal{S})}_{j,\sigma(j)} \big(\vec{p}_j \cdot \vec{\chi}_{\sigma(j)}(t_j))$$ of $N$ entries in distinct rows and columns, associated with a permutation $\sigma$, describes a probability amplitude for an $N$-photon detection \cite{Glauber2006}. Each $N$-photon amplitude depends on the interaction with the passive optical elements in the interferometer through the term $\prod_{j} \mathcal{U}^{(\mathcal{D},\mathcal{S})}_{j,\sigma(j)}$ as well as on the contribution  $\prod_{j}  \big(\vec{p}_j \cdot \vec{\chi}_{\sigma(j)}(t_j))$ depending on the photonic spectral distributions, the propagation times, the detection times and polarizations. 
The interference of all possible $N!$ detection amplitudes finally leads to the $N$-photon probability rate 
\begin{align}
	G^{(\mathcal{D},\mathcal{S})}_{\{t_j,\vec{p}_j\}}&=
\abs{\sum_{\sigma \in \Sigma_N}	\prod_{j=1}^N \mathcal{U}^{(\mathcal{D},\mathcal{S})}_{j,\sigma(j)} \big(\vec{p}_j \cdot \vec{\chi}_{\sigma(j)}(t_j))}^2\nonumber\\
	 &= \abs{\per \mathcal{T}_{\{t_j,\vec{p}_j\}}^{(\mathcal{D},\mathcal{S})}}^2
	,
	\label{eqn:CorrelationFinal}
\end{align}
uniquely defined by the permanent of the matrix in Eq.~\eqref{eqn:CorrMatrixDefintion}.
We refer to App.~\ref{app:CorrelationFunctions} for a detailed derivation of this result.

The superposition of $N!$ multi-photon amplitudes in the permanent in Eq.~\eqref{eqn:CorrelationFinal} will be fundamental in achieving computational hardness, as we will show in the next section.

\subsection{\textbf{Multi-Photon Correlation Interference in the limit \texorpdfstring{$\mathbold{N \ll M}$}{N<<M}}} \label{sec:MBCP11}

In the limit where the number $N$ of input photons is much less than the number $M$ of interferometer ports, the detection events corresponding to boson bunching can be neglected \cite{aaronson2011computational} and the port samples in Eq.~\eqref{eqn:PortSample} reduce to all the possible sets $\mathcal{D}$ of $N$ distinct values of the indices $d_j =1,2,...,M$ (see Fig. \ref{fig:InterferometerSetup}(b)).

In this case, we can directly use the indices $d\in \mathcal{D}$ and $s\in \mathcal{S}$ as labels for the detectors and the input photons, respectively, and write
\begin{align}
	G^{(\mathcal{D},\mathcal{S})}_{\{t_d, \vec{p}_d\}}
	&= \abs{\per \mathcal{T}^{(\mathcal{D},\mathcal{S})}_{\{t_d,\vec{p}_d\}}}^2, 
	\label{eqn:CorrelationFinal2}
\end{align}
with the matrices
\begin{align}
	\mathcal{T}^{(\mathcal{D},\mathcal{S})}_{\{t_d,\vec{p}_d\}} \defeq
	\big[ \mathcal{U}_{d,s} \;\big( \vec{p}_d \cdot \vec{\chi}_s(t_d) \big)
	\big]_{\substack{d\in \mathcal{D} \\ s\in \mathcal{S}}}.
		\label{eqn:CorrMatrixDefinition2}
\end{align}

\subsubsection{Occurrence of N-photon interference}
When does $N$-photon interference occur? This only happens when all the $N!$ interfering $N$-photon detection amplitudes in Eq.~\eqref{eqn:CorrelationFinal2} are non-vanishing.

\begin{figure}
	\begin{center}
			\includegraphics{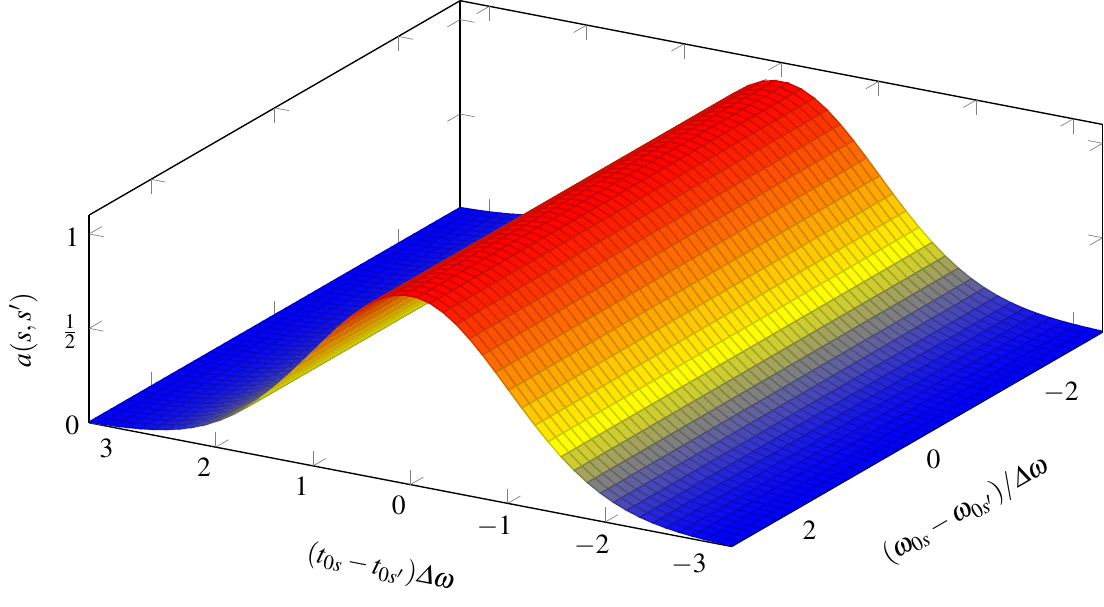}
		\caption{
	The overlap $a(s,s')$ of the moduli of the temporal distributions $\chi_s(t)$ and $\chi_{s'}(t)$ of the two photons coming from the ports $s$ and $s'$ in the case of Gaussian spectral shapes with equal bandwidth $\Delta\omega$ and equal polarization. This overlap does not depend on the central frequencies but only decays exponentially with the difference in the initial times of the photons. This is the reason why interference between photons of different colors contributes to time-resolved correlation measurements.
}
			\label{fig:OverlapModulus}
	\end{center}
\end{figure}

To establish if there is any time-polarization sample $\{t_d,\vec{p}_d\}_{d\in \mathcal{D}}$ where multi-photon interference occurs for a general interferometer transformation, we  define the $N$-photon interference matrix \cite{LaibacherTammaComplexity} with elements
\begin{align}
	a(s,s') &\defeq \int_{-\infty}^{\infty} \d{t} \abs{\vec{\chi}_{s}(t) \cdot \vec{\chi}_{s'}(t)}  
	= \abs{\vec{v}_s \cdot \vec{v}_{s'}} \int_{-\infty}^{\infty} \d{t} \abs{\vec{\chi}_s(t)} \abs{\vec{\chi}_{s'}(t)} \leq 1,
	\label{eqn:TwoPhotonOverlapModulus}
\end{align}
with $s, s' \in S$, depending on the
	pairwise overlaps of the moduli of the temporal single-photon detection amplitudes $\chi_s(t-t_{0s}-\Delta t) \ee{\ii \omega_s (t-t_{0s} - \Delta t)}$ 
and on the pairwise overlaps of the polarizations $\vec{v}_s$. The elements $a(s,s')$ are always independent of the central frequencies of the photons. For equally polarized input photons, Eq.~\eqref{eqn:TwoPhotonOverlapModulus} reduces to
\begin{align}
	a(s,s') &\defeq \int_{-\infty}^{\infty} \d{t} \abs{ \vec{\chi}_{s}(t)} \abs{ \vec{\chi}_{s'}(t)} ,
	\label{eqn:TwoPhotonOverlapModulus2}
\end{align}
where $\abs{ \vec{\chi}_{s}(t)}$ is the modulus of the vector $\vec{\chi}_{s}(t)$. As an example, for input photons with Gaussian spectral shape 
\begin{align}
	\xi_{s}(\omega) = \frac{1}{(2\pi (\Delta\omega_{s})^2)^{1/4}} \exp \left[ - \frac{\omega^2}{4(\Delta\omega_s)^2} \right]
\end{align}
with bandwidth $\Delta\omega_{s}$ in Eq.~\eqref{eqn:ComplexSpectralDistribution}, corresponding to
\begin{align}
	\chi_s(t) =  \left( \frac{2 (\Delta\omega_s)^2}{\pi} \right)^{1/4} \exp \left[ - (\Delta\omega_s)^2 t^2 \right]
\end{align}
in Eq.~\eqref{eqn:Fourier}, each interference-matrix element takes the form
\begin{align}
	a(s,s') = \abs{\vec{v}_s \cdot \vec{v}_{s'}} \sqrt{\frac{2 \Delta\omega_s \Delta\omega_{s'}}{(\Delta\omega_s)^2 + (\Delta\omega_{s'})^2}} \exp\left[ - \frac{(\Delta\omega_s)^2 (\Delta\omega_{s'})^2}{(\Delta\omega_s)^2+ (\Delta\omega_{s'})^2}(t_{0s'} - t_{0s})^2 \right],
\end{align}
which has been plotted in Fig.~\ref{fig:OverlapModulus} for $\vec{v}_s = \vec{v}_{s'}$.

In general, for non-vanishing elements
\begin{align}
	0 < a(s,s') \leq 1 \quad \forall s,s' \in \mathcal{S},
	\label{eqn:NonvanishingOverlaps}
\end{align}
a finite temporal overlap of the moduli of all single-photon detection amplitudes $ \mathcal{U}_{d,s} (\vec{e}_{\bar{\lambda}} \cdot \vec{\chi}_s(t))$ in at least one common polarization component $\vec{e}_{\bar{\lambda}}$ is ensured.
This means that a time interval $T$ and at least a polarization $\vec{e}_{\bar{\lambda}} \in \{\vec{e}_1,\vec{e}_2\}$ exist such that $\vec{e}_{\bar{\lambda}}\cdot \vec{\chi}_s(t)$ is non-vanishing for all times in $t\in T$. Therefore,  for the corresponding detection samples $\{t_d,\vec{p}_d\}_{d\in \mathcal{D}}$ with $t_d \in T$ and $\vec{p}_d = \vec{e}_{\bar{\lambda}}$,
all corresponding $N$-photon quantum-path amplitudes characterizing the permanent in Eq.~\eqref{eqn:CorrelationFinal2} are non-vanishing and contribute to the interference. Thereby, the interference matrix~\eqref{eqn:TwoPhotonOverlapModulus} represents a signature of the number of time-polarization samples $\{t_d,\vec{p}_d\}_{d\in \mathcal{D}}$ for which $N$-photon interference occurs. In Ref. \cite{LaibacherTammaComplexity}, we have demonstrated that MBCS is intractable for a classical computer with a polynomial number of resources in the condition \eqref{eqn:NonvanishingOverlaps}, assuming that the intractability of boson sampling with identical photons is correct \cite{aaronson2011computational}.

\subsubsection{N-photon quantum paths always ``indistinguishable''}\label{sec:MBCSIndist}
Further, all $N$-photon quantum paths are always indistinguishable by the detection times or polarizations if \cite{LaibacherTammaComplexity}
\begin{align}
	a(s,s') = 1 \quad \forall s,s' \in \mathcal{S}
	\label{eqn:FullTemporalOverlap}
\end{align}
in Eq.~\eqref{eqn:TwoPhotonOverlapModulus}.
Then, all $N!$ terms of the permanent in Eq.~\eqref{eqn:CorrelationFinal2} contribute to the correlation function for all possible samples $\{t_d,\vec{p}_d\}_{d\in \mathcal{D}}$. Hereafter, we obviously exclude all the trivial samples $\{t_d,\vec{p}_d\}_{d\in \mathcal{D}}$ with probability rates that vanish independently of the interferometer transformation $\mathcal{U}$.  

Complete indistinguishability by detection times or polarizations can arise in two cases. Either, all the input photons are completely identical or they differ only by their color, i.e. their central frequency. In Ref. \cite{LaibacherTammaComplexity}, we have shown how approximate MBCS with photons of different colors, even if distinguishable from each other ($\Delta\omega_{s} \ll \abs{\omega_{0s} - \omega_{0s'}} \ \forall s,s'\in \mathcal{S}$), is at least of the same complexity class as the standard boson sampling problem with identical photons.

\subsubsection{N-photon quantum paths always ``distinguishable''}\label{sec:MBCSDist}
If all input photons can be distinguished at the detectors for any possible time-polarization sample $\{t_d,\vec{p}_d\}_{d\in \mathcal{D}}$,
\begin{align}
	a(s,s') = \delta_{s,s'} \quad \forall s,s' \in \mathcal{S}
\end{align}
in Eq.~\eqref{eqn:TwoPhotonOverlapModulus}, each photon $s$ can at most contribute to the detection at one specific detector. Thus, for all possible samples $\{t_d,\vec{p}_d\}_{d\in \mathcal{D}}$, only a single $N$-photon quantum path $s \rightarrow \sigma(s)$ (with $s\in \mathcal{S}$ and a bijective map $\sigma$ from $\mathcal{S}$ to $\mathcal{D}$) contributes to the correlation function
\begin{align}
	G^{(\mathcal{D},\mathcal{S})}_{\{t_d,\vec{p}_d\}} = \Big| \prod_{s\in \mathcal{S}} \mathcal{U}_{\sigma(s),s} \big( \vec{p}_{\sigma(s)} \cdot \vec{\chi}_{s}(t_{\sigma(s)})\Big|^2.
	\label{eqn:SinglePathCorrelation}
\end{align}
Thereby, in such a case, MBCS becomes trivial from a computational point of view and, of course, this would be the same for the standard boson sampling problem.

\subsubsection{Indistinguishability at the detectors for given input photon subsets}\label{sec:MBCSDistIndist}

We now address the intermediate case where only the input photons in certain disjoint subsets $\mathcal{S}_k \subset \mathcal{S}$ of input ports, with $\bigcup_{k} \mathcal{S}_k = \mathcal{S}$, are always indistinguishable at the detectors, corresponding to
\begin{align}
	a(s,s') = 
	\begin{dcases}
		1 & s,s' \in \mathcal{S}_k \quad (\forall k)\\
		0 & s \in \mathcal{S}_k, s' \in \mathcal{S}_{k'} \quad (k\neq k').
	\end{dcases}
\end{align}
Then, for any possible time-polarization sample $\{t_d,\vec{p}_d\}_{d\in \mathcal{D}}$, multi-photon interference only occurs between $N_k <N$ photons in the same subset $\mathcal{S}_k$ with $N_k$ elements. In this case, all the possible $N$-photon detection events correspond to detector samples $\mathcal{D}$ which can be divided into subsamples $\mathcal{D}_k$ such that
\begin{align}
	\vec{p}_d \cdot \vec{\chi}_s(t_d) 
	\begin{dcases}
		\neq 0 & \forall s\in \mathcal{S}_k, d \in \mathcal{D}_k \quad (\forall k) \\
		=0 & \forall s\in \mathcal{S}_k, d \in \mathcal{D}_{k'} \quad (k\neq k').
	\end{dcases}
\end{align}
Therefore, Eq.~\eqref{eqn:CorrelationFinal2} reduces to
\begin{align}
	G^{( \mathcal{D},\mathcal{S} )}_{\{t_d,\vec{p}_d\}} = \prod_k \abs{\per \mathcal{T}^{(\mathcal{D}_k,\mathcal{S}_k)}_{\{t_d,\vec{p}_d\}_{d \in \mathcal{D}_k}}}^2,
	\label{eqn:CorrelationGroups}
\end{align}
where the matrices $\mathcal{T}^{(\mathcal{D}_k,\mathcal{S}_k)}_{\{t_d,\vec{p}_d\}_{d\in \mathcal{D}_k}}$ are now of order $N_k < N$, differently from the $N\times N$ matrices in Eq.~\eqref{eqn:CorrMatrixDefinition2}.

\section{"Non-resolved" joint detections}\label{sec:Integrated}

We now consider the case of correlation measurements which do not resolve the detection times and polarizations, resulting in an average over these degrees of freedom. In this case, we obtain the probability (see App.~\ref{app:Nonresolving:Bunching})
\begin{align}
	P_{\text{av}}(\mathcal{D};\mathcal{S}) &= \frac{1}{\prod_{d=1}^{M} n_d!} \sum_{\{\vec{p}_j\}\in \{\vec{e}_1,\vec{e}_2\}^{N}} \int_{-\infty}^{\infty} \Big( \prod_{j=1}^{N} \d{t_j} \Big) G^{(\mathcal{D},\mathcal{S})}_{\{t_j,\vec{p}_j\}}
	\label{eqn:IntegratedNotExplicit}
\end{align}
to detect the $N$ photons injected in the input ports $\mathcal{S}$ at the output ports $\mathcal{D}$, where the factor $(\prod_{d=1}^{M} n_d!)^{-1}$ arises from the symmetry of the correlation function $G^{(\mathcal{D},\mathcal{S})}_{\{t_j,\vec{p}_j\}}$ under the exchange of arguments corresponding to a detection in the same output port.

\begin{figure}
	\begin{center}
			\includegraphics{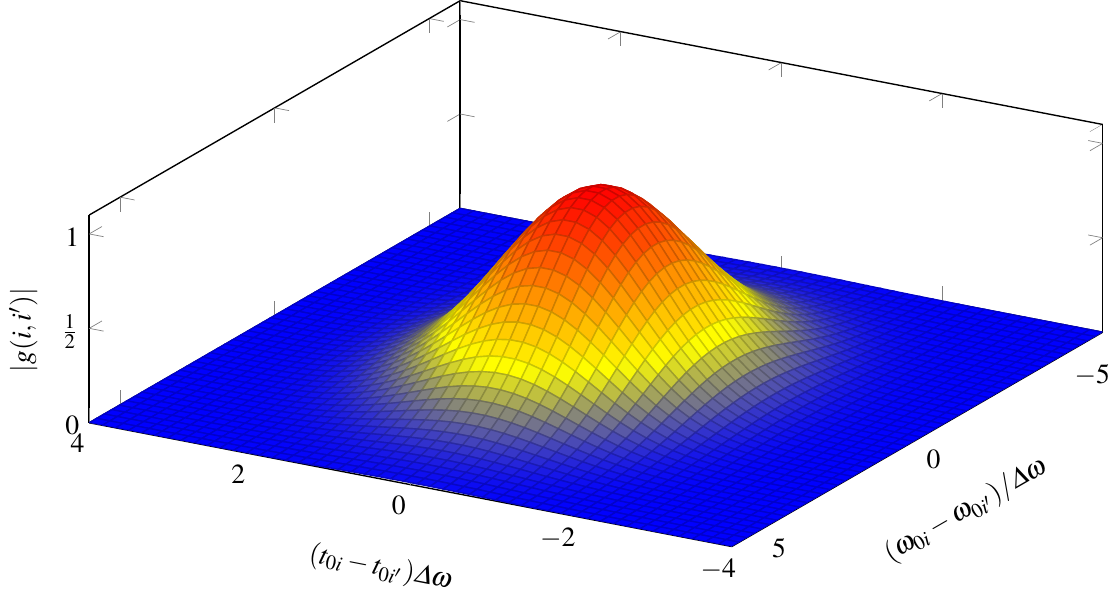}
		\caption{
				Overlap $\abs{g(i,i')}$ in Eq.~\eqref{eqn:GaussianTwoPhotonOverlap} for two photons coming from the ports $i$ and
	$i'$ in the case of Gaussian
	spectra in Eq.~\eqref{eqn:GaussianSpectrum} with equal central frequencies and equal polarizations: The overlap is maximal ($\abs{g(i,i')} = 1$) if the spectra are identical and decays exponentially with the difference $t_{0i}-t_{0i'}$ of the emission times of the photons and with the difference $\omega_{0i} - \omega_{0i'}$ of their central frequencies. 
		}
			\label{fig:Overlap}
	\end{center}
\end{figure}

In order to calculate the probability in Eq. (\ref{eqn:IntegratedNotExplicit}), it is useful to introduce the two-photon distinguishability factors \cite{TammaLaibacherPRL}
\begin{align}
g(i,i') = \int_{0}^{\infty} \d{\omega}
\vec{\xi}_i(\omega)\cdot \vec{\xi}_{i'}(\omega) = \int_{-\infty}^{\infty} \d{t} \vec{\chi}_i(t) \cdot \vec{\chi}_{i'}(t)
	\label{eqn:g},
\end{align}
with $i,i'=1,...,N$, and the \textit{$N$-photon amplitude overlaps}
\begin{align}
	f_{\rho}(\mathcal{S}) &\defeq
	\prod_{i=1}^N g(i,\rho(i)) = 
	\prod_{i=1}^N \int_{-\infty}^{\infty} \d{t} \vec{\chi}_{i}(t) \cdot \vec{\chi}_{\rho(i)}(t)
	\label{eqn:DistinguishabilityFactor}
\end{align}
with a permutation $\rho$ from the symmetric group $\Sigma_N$.
In the case of input photons with Gaussian spectral distributions 
	\begin{align}
	\vec{\xi}_{i}(\omega) = \vec{v}_i \frac{1}{(2\pi (\Delta\omega_i)^2)^{1/4}} \exp \Big( - \frac{(\omega - \omega_{0i})^2}{4 (\Delta\omega_i)^2} + \ii \omega t_{0i} \Big),
	\label{eqn:GaussianSpectrum}
\end{align}
in Eq.~\eqref{eqn:ComplexSpectralDistribution}, we find that (see App.~\ref{app:OverlapGaussianPulses})
\begin{align}
	g(i,i') &= \big( \vec{v}_i \cdot \vec{v}_{i'} \big) \sqrt{\frac{2\Delta\omega_i \Delta\omega_{i'}}{(\Delta\omega_i)^2+(\Delta\omega_{i'})^2}} \exp\left[ -\frac{(\omega_{0i}-\omega_{0i'})^2}{4((\Delta\omega_i)^2+(\Delta\omega_{i'})^2)} \right] \\
	&\hspace{3cm}\times\exp\left[ -\frac{(\Delta\omega_i)^2 (\Delta\omega_{i'})^2}{(\Delta\omega_i)^2+(\Delta\omega_{i'})^2} (t_{0i}-t_{0i'})^2\right] \\
	&\hspace{3cm} \times \exp\left[ -\ii \frac{\omega_{0i}(\Delta\omega_{i'})^2 + \omega_{0i'}(\Delta\omega_i)^2}{(\Delta\omega_i)^2 + (\Delta\omega_{i'})^2} (t_{0i}- t_{0i'}) \right].
	\label{eqn:GaussianTwoPhotonOverlap}
\end{align}
The absolute value $\abs{g(i,i')}$ is plotted in Fig.~\ref{fig:Overlap} for equal bandwidths $\Delta\omega_i = \Delta\omega_{i'}=\Delta\omega$ and equal polarizations $\vec{v}_i = \vec{v}_{i'}$.

By defining the matrices
\begin{align}
	\mathcal{A}^{(\mathcal{D},\mathcal{S})}_{\rho} &\defeq \Big[ \big( \mathcal{U}^{(\mathcal{D},\mathcal{S})}_{j,i}\big)^{*} \mathcal{U}^{(\mathcal{D},\mathcal{S})}_{j,\rho(i)} \Big]_{\substack{j=1,\dots,N \\ i=1,\dots,N}}
	\label{eqn:Amatrix},
\end{align}
the probability of an $N$-fold detection in the sample $\mathcal{D}$ can be expressed, in the narrow-bandwidth approximation, as (see App.~\ref{app:Nonresolving:Integration})
\begin{align}
	P_{\text{av}}(\mathcal{D};\mathcal{S})
	&= \frac{1}{\prod_{d=1}^{M} n_d!} \sum_{\rho\in \Sigma_N} f_{\rho}(\mathcal{S}) \per \mathcal{A}_{\rho}^{(\mathcal{D},\mathcal{S})} \\
	&= \frac{1}{\prod_{d=1}^{M} n_d!} \sum_{\rho\in \SymmGroup{\Ndets}}
		\OverlapFactor{\rho} \sum_{\permut\in\Sigma_N} \left[
			\prod_{j=1}^N \big( \mathcal{U}^{(\mathcal{D},\mathcal{S})}_{j,\sigma(j)} \big)^{*}
			\mathcal{U}^{(\mathcal{D},\mathcal{S})}_{j,\rho(\sigma(j))}
								 \right].
		\label{eqn:IntegratedGeneral} 
\end{align}

The time- and polarization-averaged probability $\TotalProb{\BSPindex}$ in
Eq.~\eqref{eqn:IntegratedGeneral} associated with the detection of $\Ndets$ photons in the
$\Ndets$-port sample $\setD$ comprises $\Ndets!$ contributions for each
permutation $\rho\in\SymmGroup{\Ndets}$. 
Each contribution contains all $\Ndets!$ cross terms $\prod_{j=1}^N \big( \mathcal{U}^{(\mathcal{D},\mathcal{S})}_{j,\sigma(j)} \big)^{*} \mathcal{U}^{(\mathcal{D},\mathcal{S})}_{j,\rho(\sigma(j))}$
arising from the interference of the interferometer-dependent multi-photon
amplitudes $\prod_{j=1}^{N} \mathcal{U}^{(\mathcal{D},\mathcal{S})}_{j,\sigma(j)}$, with
$\sigma \in \Sigma_N$, in the condition that the $\Ndets$ photon pairs
$\left\{(i,\rho(i))\right\}_{i=1,\dots,N}$ for each cross
term are fixed by a given permutation $\rho$. Moreover, each factor
$\OverlapFactor{\rho}$
describes the degree of pairwise indistinguishability for the set $\left\{(i,\rho(i))\right\}_{i=1,\dots,N}$ of source pairs. 

\begin{figure}
	\begin{center}
		\includegraphics[scale=0.98]{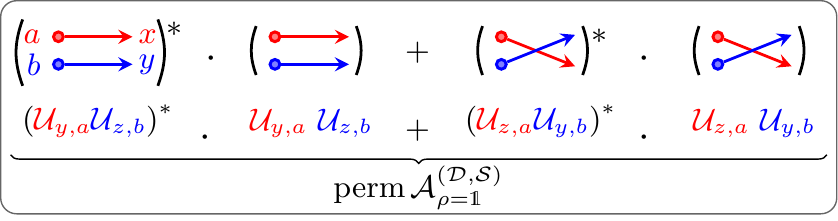} 
		\includegraphics[scale=0.98]{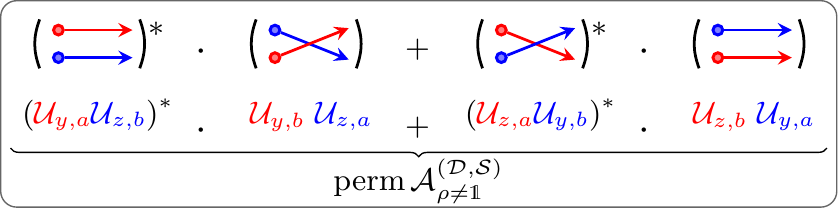} 
		\caption{Physical illustration of the permanent terms in
			Eq.~\eqref{eqn:Nonresolving} for $\Ndets=2$ ($\setS=\left\{ a,b
			\right\}$,$\setD=\left\{ y,z \right\}$) in the two possible cases
			$\rho=\unity$ (indistinguishability factor
			$\OverlapFactor{\rho=\unity}=1$) and $\rho\neq\unity$
			($\OverlapFactor{\rho\neq\unity}\neq 1$ for incomplete overlap of the
			photonic spectral distributions). In both cases,
			$\Ndets=2$ pairs of sources $\left\{ (i,\rho(i))
			\right\}_{i=1,2}$, each indicated by a separate color, are
			coupled in the $\Ndets!=2$ possible interference terms defined by the
			two ways of connecting the two pairs to the two detectors $y$ and
			$z$. These possibilities sum up to the permanents $\per
						\mathcal{A}^{(\setD,\setS)}_{\rho} $. Thereby,
						Eq.~\eqref{eqn:IntegratedGeneral} becomes
						$\TotalProb{\BSPindex}= \abs{\Uto{a}{y}\Uto{b}{z}}^2 +
						\abs{\Uto{b}{y}\Uto{a}{z}}^2 +
						\abs{\OverlapFrequency{1}{2}}^2 2 \Re\left(
						\Utoconj{a}{y}\Utoconj{b}{z} \Uto{b}{y}\Uto{a}{z} \right)$.
}
		\label{fig:InterferenceTerms}
	\end{center}
\end{figure}

\subsection{\textbf{``Non-resolved'' multi-boson interference in the limit \texorpdfstring{$\mathbold{N \ll M}$}{N<<M}}} \label{sec:Integrated1}
As pointed out in section~\ref{sec:MBCP11}, in this limit the port samples in Eq.~\eqref{eqn:PortSample} reduce to all the possible sets $\mathcal{D}$ of $N$ distinct values of the index $d =1,2,...,M$. 
Thereby, since $n_d=1 \ \forall d\in \mathcal{D}$ and $n_d = 0 \ \forall d \notin \mathcal{D}$, Eq.~\eqref{eqn:IntegratedGeneral} becomes \cite{TammaLaibacherPRL}
\begin{align}
	P_{\text{av}}(\mathcal{D};\mathcal{S})
	= \sum_{\rho \in \Sigma_N} f_{\rho}(\mathcal{S}) \per \mathcal{A}_{\rho}^{(\mathcal{D},\mathcal{S})}.
	\label{eqn:Nonresolving}
\end{align}
Alternate expressions for the detection probabilities in Eq.~\eqref{eqn:Nonresolving} can also be found in \cite{Rohde2015,Tichy2015,Shchesnovich2015}.

In Fig.~\ref{fig:InterferenceTerms} we illustrate as an example the case of
$\Ndets=2$.
Here, for a unitary matrix $\mathcal{U}$ describing a 50/50 beam splitter transformation, Eq.~\eqref{eqn:Nonresolving} fully describes the famous two-photon interference dip \cite{Hong1987,Alley1986,Tamma2015}.

Interestingly, for partially distinguishable photons, the complexity of the problem of sampling from the probability distribution in Eq.~\eqref{eqn:Nonresolving} is still an open question \cite{Tamma2015}. Here, we will address the two limiting cases of identical and fully distinguishable input photons.

\subsubsection{Identical input photons: Boson Sampling} \label{sec:BSP}
In the case of identical input photons, all photonic spectral distributions fully overlap pairwisely, corresponding to the complete overlaps $$\OverlapFactor{\rho}\approx 1 \ \forall \rho\in \Sigma_N$$ associated with the interfering $N$-photon amplitudes.
Accordingly, Eq.~\eqref{eqn:Nonresolving} reduces to \cite{TammaLaibacherPRL}
\begin{align}
	\TotalProb{\text{av}} & \approx 
	\sum_{\rho \in \SymmGroup{\Ndets}}^{} \per \ProdMatrix{\rho}{}
	= \abs{\per \Umatrix{}}^2.
	\label{eqn:IdenticalPhots}
\end{align}
This dependence of all the probabilities $\TotalProb{\BSPindex}$ on  permanents of random complex matrices, given by submatrices of the random unitary matrix $\mathcal{U}$, is at the heart of the computational complexity of the boson sampling problem \cite{aaronson2011computational}.

\subsubsection{Distinguishable input photons} \label{sec:Disting}
This case corresponds to $N$-photon amplitude overlaps $$f_{\rho}(\mathcal{S}) \approx 0 \ \forall \rho\neq\unity,$$ leading to no multi-photon interference.  The probability in
		Eq.~\eqref{eqn:IntegratedGeneral} is given by the completely
		incoherent superposition \cite{TammaLaibacherPRL}
\begin{align}
	P_{\text{av}}(\mathcal{D};\mathcal{S}) &\approx \per \mathcal{A}^{(\mathcal{D},\mathcal{S})}_{\rho=\unity}
	\label{Pind}
\end{align}
with the non-negative matrix
$\mathcal{A}^{(\mathcal{D},\mathcal{S})}_{\rho=\unity} = [ \abs{\mathcal{U}_{d,s}}^2 ]_{\substack{d\in \mathcal{D} \\ s\in \mathcal{S}}}$,
whose permanent can be efficiently estimated \cite{Jerrum2004}.

\section{Discussion}
We provided a full analysis of the MBCS problem, based on time- and polarization-resolved multi-boson correlation measurements in passive, linear optical networks. For simplicity, we assumed approximately equal propagation times for all possible paths from the sources to the detectors. We derived compact expressions in terms of permanents for the detection probabilities for input bosons with arbitrary spectral distributions, including the case of bunching of the bosons at the output of the interferometer.
Interestingly, a permanent structure already emerges at an operator level and holds for a broader range of states \cite{TammaLaibacher2} than just the single-photon states considered here. This permanent structure leads to detection probability rates given by the modulus square of matrix permanents depending on the interferometer evolution as well as on the detection times and polarizations (see Eq.~\eqref{eqn:CorrelationFinal2}).

To get a deeper physical understanding of the connection between the state of the input photons, the occurrence of multi-photon interference, and the complexity of MBCS, we discussed three limiting cases which differ from each other in the highest order of multi-photon interference that can be observed and, thereby, in their inherent complexity.

For single photons impinging on the interferometer at distinct times with respect to their coherence times, the moduli of the temporal single-photon detection amplitudes do not overlap pairwisely. Since this allows us to determine which input photon causes a given detector ``click'', the photons cannot interfere at all and only a single multi-photon quantum path contributes to an $N$-photon detection event (see Eq.~\eqref{eqn:SinglePathCorrelation}). Obviously, the MBCS problem is computationally trivial in this case. 

We also addressed the case where the input photons can be divided into temporally distinguishable subsets, such that only photons of the same subset are pairwisely indistinguishable in time. In this case, interference only occurs between photons of the same subset. The detection probabilities are then given by the product of the respective detection probabilities for each subset exhibiting interference of a reduced number $N_k < N$ of photons (see~Eq.~\eqref{eqn:CorrelationGroups}). 

Differently, detection events in which all $N$ photons interfere correspond to non-vanishing contributions of all $N!$ $N$-photon amplitudes in Eq.~\eqref{eqn:CorrelationFinal2}.
This can only occur if the photons share a common polarization component and if all pairwise overlaps of the absolute values of the temporal single-photon detection amplitudes are non-vanishing in that polarization (see condition~\eqref{eqn:NonvanishingOverlaps}). 
Further, if the moduli of these temporal amplitudes and the polarizations are all identical,
all possible $N$-photon detections correspond to $N$-photon interference events.
Surprisingly, this occurs also for input states with photons of different colors since the moduli of the temporal single-photon detection amplitudes are independent of the photon colors. Therefore, full $N$-boson interference is possible even for photons which are fully distinguishable in their colors and thus in their spectral distributions (vanishing pairwise indistinguishability factors in Eq.~\eqref{eqn:g}). 

The reason that $N$-photon interference is observed despite the color differences of the photons lies in the extremely short integration times of the detectors with respect to the inverse of the differences of the photon frequencies. Since resolutions in time and frequency are reciprocal to each other, these detectors cannot resolve 
the differences in the colors of the photons, leading to an effective indistinguishability of the photons.
This highlights the fact that the observation of interference does not require the particles to be identical but only requires them to ``appear identical'' in a given correlation measurement.
However, the visibility of the $N$-photon interference, which occurs independently of the photon colors in multi-photon correlation measurements (see Fig.~\ref{fig:OverlapModulus}), decreases exponentially with respect to the color differences when the observer ignores the information about the detection times (see Fig.~\ref{fig:Overlap}).

From a computational point of view, multi-photon interference events of highest order $N$ are at the heart of the complexity of MBCS. Indeed, the classical intractability of exact MBCS \cite{LaibacherTammaComplexity} relies on the occurrence of such events (guaranteed by condition~\eqref{eqn:NonvanishingOverlaps}).
Interestingly, when \textit{all} possible detection events exhibit $N$-photon interference even approximate MBCS is classically intractable with a polynomial number of resources \cite{LaibacherTammaComplexity}. This is important from an experimental point of view, where the inevitable experimental errors prevent exact sampling. As already pointed out, the hardness of approximate MBCS even holds for photons of completely different colors.
This remarkable result proves that the complexity of multi-boson correlation measurements is actually determined by the effective indistinguishability of the photons at the detectors ensuring full $N$-photon interference.
In contrast, by classically averaging over the detection times and polarizations as in the standard boson sampling problem, the information about such full $N$-photon interference and its inherent computational power is lost (see Eq. \eqref{Pind}).
Here, while the MBCS problem is classically intractable, the corresponding boson sampling problem becomes trivial \cite{LaibacherTammaComplexity}. The extension of the analysis of the approximate MBCS problem to photons with arbitrary temporal distributions at the detectors will be addressed in a further publication.

In conclusion, our results demonstrate the surprising computational power of multi-boson correlation interferometry even with input bosons which are partially or fully distinguishable.
This represents a significant step forward towards the implementation of ``real-world'' interferometric networks in quantum information processing \cite{Tamma2014,Tamm2015a,Tamm2015,Tamma2009,tamma2011factoring,Tamma2012} overcoming the experimental need for identical bosons.

\section*{Acknowledgments}
V.T. is extremely glad to dedicate this work to the memory of Professor Howard Brandt. His enormous contribution to the field of quantum information is a treasure which will benefit science forever.

\section*{Funding}
V.T. acknowledges the support of the German Space Agency DLR with funds provided by the Federal Ministry of Economics and Technology (BMWi) under grant no. DLR 50 WM 1556.
This work was supported by a grant from the Ministry of Science, Research and the Arts of Baden-W\"urttemberg (Az: 33-7533-30-10/19/2).

\appendix

\section{Correlation functions as expectation values of permanents of operator matrices}\label{app:OperatorPermanents}
We derive the expression in Eq.~\eqref{eqn:CorrelationOperatorMatrix} of the correlation function $G^{(\mathcal{D},\mathcal{S})}_{\{t_j,\vec{p}_j\}}$ for multimode single photon states.
Inserting Eq.~\eqref{eqn:ExpansionAout} into Eq.~\eqref{eqn:ProbabilityRate}, we obtain
\begin{align}
	G^{(\mathcal{D},\mathcal{S})}_{\{t_j,\vec{p}_j\}} = \matrixel{\mathcal{S}}{\prod_{j=1}^N \Big[ \sum_{i'=1}^N \big( \mathcal{U}^{(\mathcal{D},\mathcal{S})}_{j,i'}\big)^{*} \Big( \vec{p}_j^{*} \cdot \hat{\vec{E}}_{i'}^{(-)}(t_j -\Delta t) \Big) \Big] \Big[ \sum_{i=1}^N \mathcal{U}^{(\mathcal{D},\mathcal{S})}_{j,i} \Big( \vec{p}_j \cdot \hat{\vec{E}}_{i}^{(+)}(t_j -\Delta t) \Big) \Big] }{\mathcal{S}},
	\intertext{which, after interchanging the order of the product and the two summations, becomes}
	G^{(\mathcal{D},\mathcal{S})}_{\{t_j,\vec{p}_j\}} = \sum_{ \{ i_j,  i_j' \}_{j=1,\dots,N} }\,\, \prod_{j=1}^N \big( \mathcal{U}^{(\mathcal{D},\mathcal{S})}_{j,i_j'} \big)^{*} \mathcal{U}^{(\mathcal{D},\mathcal{S})}_{j,i_j} \matrixel{\mathcal{S}}{\prod_{j=1}^N \Big( \vec{p}^{*}_j \cdot \hat{\vec{E}}^{(-)}_{i_j'}(t_j - \Delta t) \Big) \prod_{j=1}^N \Big( \vec{p}_j \cdot \hat{\vec{E}}^{(+)}_{i_j}(t_j - \Delta t) \Big)}{\mathcal{S}}.
	\label{eqn:ProbRateStart}
\end{align}
Here, the summation over the $2N$ indices $\{i_j,i_j'\}_{j=1,\dots,N}$ covers all possible ways the $N$ sources contribute to the product of $2N$ field operators in the expectation value in Eq.~\eqref{eqn:ProbabilityRate}, independently of the input state.

Eq.~\eqref{eqn:ProbRateStart} can be simplified by recalling that the input state is the product of single photon states in each of the input ports. Thus, since each source can at most contribute one photon, the expression $\prod_{j=1}^N \big( \vec{p}_j \cdot \hat{\vec{E}}^{(+)}_{i_j}(t_j-\Delta t) \big) \ket{\mathcal{S}}$ is non-vanishing only if the indices $i_j\ (j=1,\dots,N)$ take pairwise different values.
Introducing the symmetric group $\Sigma_N$ of all permutations $\sigma$ of $N$ elements, we find that in the non-vanishing cases, the indices take the values $i_j = \sigma(j)$, with $\sigma \in \Sigma_N$. The same argument can also be made for the possible values of the indices $i_j'$.
Therefore, the correlation function in Eq.~\eqref{eqn:ProbRateStart} reduces to
\begin{align}
	G^{(\mathcal{D},\mathcal{S})}_{\{t_j,\vec{p}_j\}} &=
	\sum_{\sigma,\sigma'\in \Sigma_N} \prod_{j=1}^N
	\big( \mathcal{U}^{(\mathcal{D},\mathcal{S})}_{j,\sigma'(j)} \big)^{*} \mathcal{U}^{(\mathcal{D},\mathcal{S})}_{j,\sigma(j)}	\matrixel{\mathcal{S}}{\prod_{j=1}^N \Big( \vec{p}^{*}_j \cdot \hat{\vec{E}}^{(-)}_{\sigma'(j)}(t_j - \Delta t) \Big)
	\Big( \vec{p}_j \cdot \hat{\vec{E}}^{(+)}_{\sigma(j)}(t_j - \Delta t) \Big)}{\mathcal{S}}, \hspace{0.5cm}
	\label{eqn:SimplifiedCorrelationFunction}
\end{align}
which, with the definition of the operator matrix $\hat{\mathcal{M}}_{\{t_j,\vec{p}_j\}}^{(\mathcal{D},\mathcal{S})}$ in Eq.~\eqref{eqn:OperatorMatrixDef}, becomes
\begin{align}
	G^{(\mathcal{D},\mathcal{S})}_{\{t_j,\vec{p}_j\}} &=
	\matrixel{\mathcal{S}}{
		\Big[ \sum_{\sigma'\in \Sigma_N} \prod_{j=1}^N \big( \mathcal{U}^{(\mathcal{D},\mathcal{S})}_{j,\sigma'(j)} \big)^{*} \Big( \vec{p}^{*}_j \cdot \hat{\vec{E}}^{(-)}_{\sigma'(j)}(t_j - \Delta t) \Big) \Big]
		\Big[ \sum_{\sigma\in \Sigma_N} \prod_{j=1}^N \mathcal{U}^{(\mathcal{D},\mathcal{S})}_{j,\sigma(j)} \Big( \vec{p}_j \cdot \hat{\vec{E}}^{(+)}_{\sigma(j)}(t_j - \Delta t) \Big) \Big]
}{\mathcal{S}}  \\
&= \matrixel{\mathcal{S}}{\Big(\per \hat{\mathcal{M}}^{(\mathcal{D},\mathcal{S})}_{\{t_j,\vec{p}_j\}}\Big)^{\dagger} \Big( \per \hat{\mathcal{M}}^{(\mathcal{D},\mathcal{S})}_{\{t_j,\vec{p}_j\}} \Big)}{\mathcal{S}}.
\end{align}

\section{Correlation functions for multimode single photon states}\label{app:CorrelationFunctions}

We demonstrate here the expression for the correlation function for multimode single photon states, found in Eq.~\eqref{eqn:CorrelationFinal}. From the definition of the input state in Eq.~\eqref{eqn:StateDefinition}, we find the expression 
\begin{align}
	\prod_{j=1}^N \Big( \vec{p}_j \cdot \hat{\vec{E}}^{(+)}_{\sigma(j)} (t_j - \Delta t) \Big) \ket{\mathcal{S}}  
	&= \prod_{j=1}^N \Big( \vec{p}_j \cdot \hat{\vec{E}}^{(+)}_{\sigma(j)} (t_j - \Delta t) \Big) \bigotimes_{i=1}^N \ket{1[\vec{\xi}_i]}_{s_i} \bigotimes_{s\notin \mathcal{S}} \ket{0}_s  \\
	&= \bigotimes_{j=1}^N \Big[ \Big( \vec{p}_j \cdot \hat{\vec{E}}^{(+)}_{\sigma(j)} (t_j - \Delta t) \Big) \ket{1[\vec{\xi}_{\sigma(j)}]}_{s_{\sigma(j)}} \Big] \bigotimes_{s\notin \mathcal{S}} \ket{0}_s 
	\label{eqn:ProdOfStates}
\end{align}
for the terms in Eq.~\eqref{eqn:SimplifiedCorrelationFunction}.
By using the definition of the single photon states in Eq.~\eqref{eqn:SinglePhotonState} and the field operators
\begin{align}
	\hat{\vec{E}}^{(+)}_i (t) = \frac{1}{\sqrt{2\pi}} \sum_{\lambda=1,2} \vec{e}_{\lambda}  \int_0^{\infty} \d{\omega}\,\ee{-\ii \omega t} \hat{a}_{i,\lambda}(\omega)
\end{align}
in the narrow-bandwidth approximation, we obtain
\begin{align}
	\Big( \vec{p}_j \cdot \hat{\vec{E}}^{(+)}_i (t_j- \Delta t) \Big) \ket{1[\vec{\xi}_i]}_{s_i} &= \frac{1}{\sqrt{2\pi}} \sum_{\lambda=1,2} (\vec{p}_j \cdot \vec{e}_{\lambda}) \int_0^{\infty} \d{\omega} \ee{-\ii \omega (t_j-\Delta t)} \hat{a}_{i,\lambda}(\omega) \sum_{\lambda'=1,2} \int_0^{\infty} \d{\omega'} ( \vec{e}_{\lambda'} \cdot \vec{\xi}_{i}(\omega') ) \hat{a}^{\dagger}_{i,\lambda'}(\omega') \ket{0}_{s_i} \\
	&\hspace{-2cm}= \frac{1}{\sqrt{2\pi}} \sum_{\lambda,\lambda'=1,2} \int_0^{\infty} \d{\omega} \int_0^{\infty} \d{\omega'} (\vec{p}_j \cdot \vec{e}_{\lambda}) \ee{-\ii \omega (t_j-\Delta t)}\big( \vec{e}_{\lambda'} \cdot \vec{\xi}_{i}(\omega') \big)  \hat{a}_{i,\lambda}(\omega) \hat{a}^{\dagger}_{i,\lambda'}(\omega') \ket{0}_{s_i} \\
	&\hspace{-2cm}= \frac{1}{\sqrt{2\pi}} \sum_{\lambda,\lambda'=1,2} \int_0^{\infty} \d{\omega} \int_0^{\infty} \d{\omega'} (\vec{p}_j \cdot \vec{e}_{\lambda}) \ee{-\ii \omega (t_j-\Delta t)}\big( \vec{e}_{\lambda'} \cdot \vec{\xi}_{i}(\omega') \big) \big[  \hat{a}^{\dagger}_{i,\lambda'}(\omega')\hat{a}_{i,\lambda}(\omega) + \delta_{\lambda,\lambda'} \delta(\omega- \omega') \big] \ket{0}_{s_i} \\
	&\hspace{-2cm}= \frac{1}{\sqrt{2\pi}} \sum_{\lambda=1,2} \int_0^{\infty} \d{\omega} (\vec{p}_j \cdot \vec{e}_{\lambda}) \ee{-\ii \omega (t_j-\Delta t)}\big( \vec{e}_{\lambda} \cdot \vec{\xi}_{i}(\omega) \big) \ket{0}_{s_i} \\
	&\hspace{-2cm}= \frac{1}{\sqrt{2\pi}} \int_0^{\infty} \d{\omega} \big(\vec{p}_j \cdot \vec{\xi}_i(\omega)\big) \ee{-\ii \omega (t_j-\Delta t)} \ket{0}_{s_i}.
\end{align}
In the narrow bandwidth approximation, we can approximately expand the integration over $\omega$ to the complete real axis. By using the definition of a Fourier transform
\begin{align}
	\mathcal{F}[f](t) \defeq \frac{1}{\sqrt{2\pi}} \int_{-\infty}^{\infty} \d{\omega} f(\omega) \ee{-\ii \omega t}
\end{align}
we then find
\begin{align}
	\Big( \vec{p}_j \cdot \hat{\vec{E}}^{(+)}_i (t_j- \Delta t) \Big) \ket{1[\vec{\xi}_i]}_{s_i} 
	&\approx \frac{1}{\sqrt{2\pi}} \int_{-\infty}^{\infty} \d{\omega} \big(\vec{p}_j \cdot \vec{\xi}_i(\omega)\big) \ee{-\ii \omega (t_j-\Delta t)} \ket{0}_{s_i} \\
	&= \big( \vec{p}_j \cdot \mathcal{F}[\vec{\xi}_i](t_j-\Delta t) \big) \ket{0}_{s_i} = \big( \vec{p}_j \cdot \vec{\chi}_s(t_j) \big) \ket{0}_{s_i}.
\end{align}
By substituting this result together with Eq.~\eqref{eqn:ProdOfStates}, Eq.~\eqref{eqn:SimplifiedCorrelationFunction} reduces to the expression
\begin{align}
	G^{(\mathcal{D},\mathcal{S})}_{\{t_j,\vec{p}_j\}} 
	&= \sum_{\sigma,\sigma'\in\Sigma_N} \prod_{j=1}^N \big( \mathcal{U}^{(\mathcal{D},\mathcal{S})}_{j,\sigma'(j)} \big)^{*} \mathcal{U}^{(\mathcal{D},\mathcal{S})}_{j,\sigma(j)} 
	\big( \vec{p}_j \cdot \vec{\chi}_{\sigma'(j)}(t_j) \big)^{*} \big( \vec{p}_j \cdot \vec{\chi}_{\sigma(j)}(t_j) \big) 
	\label{eqn:Supp:ExplicitTimeResolving}
	\\
	&= \Big\lvert \sum_{\sigma \in \Sigma_N} \prod_{j=1}^N \mathcal{U}^{(\mathcal{D},\mathcal{S})}_{j,\sigma(j)} \big(\vec{p}_j \cdot \vec{\chi}_{\sigma(j)}(t_j) \big) \Big\rvert^2 = \abs{\per \mathcal{T}^{(\mathcal{D},\mathcal{S})}_{\{t_j,\vec{p}_j\}}}^2
	\label{eqn:Supp:RateAppendix}
\end{align}
in Eq.~\eqref{eqn:CorrelationFinal}.

\section{Probabilities for non-resolved detections}\label{app:Nonresolving}

\subsection{Accounting for photon bunching}\label{app:Nonresolving:Bunching}
We now derive the probabilities~\eqref{eqn:IntegratedNotExplicit} in the case of non-resolved detections in time and polarization. Towards this end, we have to take into account that the correlation function in Eq.~\eqref{eqn:ProbabilityRate} is symmetric under permutation of the arguments $\{t_j,\vec{p}_j\}$ and $\{t_{j'},\vec{p}_{j'}\}$ if both corresponding photons are detected in the same output port, i.e. if $d_j = d_{j'}$.

For example, in the case of $N=2$ and with both photons detected in the same output port $d$ ($\mathcal{D}=\{d,d\}$), the correlation function
\begin{align}
	G^{(\{d,d\},\mathcal{S})}_{\{t_1,\vec{p}_1;t_2,\vec{p}_2\}} = \matrixel{\mathcal{S}}{\Big( \vec{p}^{*}_1 \cdot \hat{\vec{E}}^{(-)}_d(t_1) \Big)\Big( \vec{p}^{*}_2 \cdot \hat{\vec{E}}^{(-)}_d(t_2) \Big)\Big( \vec{p}_2 \cdot \hat{\vec{E}}^{(+)}_d(t_2) \Big)\Big( \vec{p}_1 \cdot \hat{\vec{E}}^{(+)}_d(t_1) \Big)}{\mathcal{S}} 
\end{align}
is symmetric with respect to the two possible detection times and respective polarizations $\{t_1,\vec{p}_1\}$ and $\{t_2,\vec{p}_2\}$: $G^{(\{d,d\},\mathcal{S})}_{\{t_1,\vec{p}_1;t_2,\vec{p}_2\}} = G^{(\{d,d\},\mathcal{S})}_{\{t_2,\vec{p}_2;t_1,\vec{p}_1\}}$. This reflects the fact that both expressions describe the same 
probability rate of two photons being detected in the output port $d$, one at time $t_1$ and with polarization $\vec{p}_1$ and the other at time $t_2$ and with polarization $\vec{p}_2$.

Therefore, the corresponding probability
\begin{align}
	P_{\text{av}}(\{d,d\};\mathcal{S}) =\frac{1}{2!} \sum_{\vec{p}_1,\vec{p}_2 \in \{\vec{e}_1,\vec{e}_2\}}\int_{-\infty}^{\infty} \d{t_1} \int_{-\infty}^{\infty}\d{t_2} G^{(\mathcal{D},\mathcal{S})}_{\{t_1,\vec{p}_1;t_2,\vec{p}_2\}}
\end{align}
for non-resolved detections contains the additional factor $1/2!$ in order to avoid double counting.

Generalizing to the case of arbitrary samples $\mathcal{D}$ in a $2M$-port interferometer, where $d$ is contained $n_d$ times in $\mathcal{D}$ ($\sum_{d=1}^{M} n_d = N$), we find the time- and polarization-averaged probabilities
\begin{align}
	P_{\text{av}}(\mathcal{D};\mathcal{S}) &= \frac{1}{\prod_{d=1}^{M} n_d!} \sum_{\{\vec{p}_j\}\in \{\vec{e}_1,\vec{e}_2\}^{N}} \int_{-\infty}^{\infty} \Big( \prod_{j=1}^{N} \d{t_j} \Big) G^{(\mathcal{D},\mathcal{S})}_{\{t_j,\vec{p}_j\}} 
	\label{eqn:Supp:Integrated}
\end{align}
in Eq.~\eqref{eqn:IntegratedNotExplicit}.

\subsection{Integrated probabilities}\label{app:Nonresolving:Integration}
We derive now Eq.~\eqref{eqn:IntegratedGeneral} from the general expression of the probability $P_{\text{av}}(\mathcal{D};\mathcal{S})$ in Eq.~\eqref{eqn:IntegratedNotExplicit} for non-resolved detections.
We first substitute in Eq.~\eqref{eqn:IntegratedNotExplicit} the value for the correlation function in Eq.~\eqref{eqn:Supp:ExplicitTimeResolving}, yielding
\begin{align}
	P_{\text{av}}(\mathcal{D};\mathcal{S}) &= \\
	& \hspace{-1cm}\frac{1}{\prod_{d=1}^{M} n_d!} \sum_{\{\vec{p}_j\}\in\{\vec{e}_1,\vec{e}_2\}^{N}} \int_{-\infty}^{\infty}\Big( \prod_{j=1}^N \d{t_j} \Big) \sum_{\sigma,\sigma'\in \Sigma_N} \prod_{j=1}^N \big( \mathcal{U}^{(\mathcal{D},\mathcal{S})}_{j,\sigma(j)} \big)^{*} \mathcal{U}^{(\mathcal{D},\mathcal{S})}_{j,\sigma'(j)}\big(\vec{p}_j \cdot \vec{\chi}_{\sigma(j)}(t_j) \big)^{*} \big( \vec{p}_j \cdot \vec{\chi}_{\sigma'(j)}(t_j) \big) \\
	& = \frac{1}{\prod_{d=1}^{M} n_d!} \sum_{\sigma,\sigma'\in \Sigma_N} \prod_{j=1}^N \big( \mathcal{U}^{(\mathcal{D},\mathcal{S})}_{j,\sigma(j)} \big)^{*} \mathcal{U}^{(\mathcal{D},\mathcal{S})}_{j,\sigma'(j)} \Bigg[ \sum_{\vec{p}_j \in \{\vec{e}_1, \vec{e}_2\}} \int_{-\infty}^{\infty} \d{t_j} \big(\vec{p}_j \cdot \vec{\chi}_{\sigma(j)}(t_j) \big)^{*} \big( \vec{p}_j \cdot \vec{\chi}_{\sigma'(j)}(t_j) \big) \Bigg] \\
	& = \frac{1}{\prod_{d=1}^{M} n_d!} \sum_{\sigma,\sigma'\in \Sigma_N} \prod_{j=1}^N \big( \mathcal{U}^{(\mathcal{D},\mathcal{S})}_{j,\sigma(j)} \big)^{*} \mathcal{U}^{(\mathcal{D},\mathcal{S})}_{j,\sigma'(j)} \Bigg[ \int_{-\infty}^{\infty} \d{t} \vec{\chi}_{\sigma(j)}(t) \cdot \vec{\chi}_{\sigma'(j)}(t)\Bigg].
	\label{eqn:supp:bintermediate}
\end{align}
With the substitution $\sigma' \rightarrow \rho \circ \sigma$, where $\rho \in \Sigma_N$, Eq.~\eqref{eqn:supp:bintermediate} becomes
\begin{align}
	P_{\text{av}}(\mathcal{D};\mathcal{S}) = \frac{1}{\prod_{d=1}^{M} n_d!} \sum_{\rho\in \Sigma_N} \sum_{\sigma \in \Sigma_N} \Bigg[\prod_{j=1}^N \big( \mathcal{U}^{(\mathcal{D},\mathcal{S})}_{j,\sigma(j)} \big)^{*} \mathcal{U}^{(\mathcal{D},\mathcal{S})}_{j,\rho(\sigma(j))} \Bigg] \Bigg[ \prod_{j=1}^N \int_{-\infty}^{\infty} \d{t} \vec{\chi}_{\sigma(j)}(t) \cdot \vec{\chi}_{\rho(\sigma(j))}(t) \Bigg].
\end{align}
Here, the second product can be written as a product over $i=\sigma(j)$ instead of $j$ since $\sigma$ is bijective, rendering this expression independent of the permutation $\sigma$:
\begin{align}
	P_{\text{av}}(\mathcal{D};\mathcal{S}) = \frac{1}{\prod_{d=1}^{M} n_d!} \sum_{\rho\in \Sigma_N} \Bigg[ \prod_{i=1}^N \int_{-\infty}^{\infty} \d{t} \vec{\chi}_{i}(t) \cdot \vec{\chi}_{\rho(i)}(t) \Bigg] \sum_{\sigma \in \Sigma_N} \Bigg[\prod_{j=1}^N \big( \mathcal{U}^{(\mathcal{D},\mathcal{S})}_{j,\sigma(j)} \big)^{*} \mathcal{U}^{(\mathcal{D},\mathcal{S})}_{j,\rho(\sigma(j))} \Bigg].
	\label{eqn:Supp:SecondLastIntegrated}
\end{align}
By using the definitions of $f_{\rho}(\mathcal{S})$ and $\mathcal{A}_{\rho}^{(\mathcal{D},\mathcal{S})}$ in Eqs.~\eqref{eqn:DistinguishabilityFactor} and \eqref{eqn:Amatrix},  Eq.~\eqref{eqn:Supp:SecondLastIntegrated} finally reduces to the expression
\begin{align}
	P_{\text{av}}(\mathcal{D};\mathcal{S}) = \frac{1}{\prod_{d=1}^{M} n_d!} \sum_{\rho\in \Sigma_N} f_{\rho}(\mathcal{S}) \per A_{\rho}^{(\mathcal{D},\mathcal{S})}
\end{align}
in Eq.~\eqref{eqn:IntegratedGeneral}.

\section{Two-photon indistinguishability factors for Gaussian spectral distributions}\label{app:OverlapGaussianPulses}

We derive the two-photon indistinguishability factors for Gaussian spectral distributions in Eq.~\eqref{eqn:GaussianTwoPhotonOverlap}.
By inserting Eq.~\eqref{eqn:GaussianSpectrum} into Eq.~\eqref{eqn:g}, we obtain the Gaussian integral
\begin{align}
	g(i,i') = \big( \vec{v}_i \cdot \vec{v}_{i'} \big) \frac{1}{(2\pi \Delta\omega_i\Delta\omega_{i'})^{1/2}} \int_{-\infty}^{\infty} \d{\omega} \exp\left[ -\frac{(\omega-\omega_{0i})^2}{4(\Delta\omega_i)^2} - \ii \omega \, t_{0i} \right] \exp\left[ -\frac{(\omega-\omega_{0i'})^2}{4(\Delta\omega_{i'})^2} + \ii \omega\, t_{0i'} \right],
\end{align}
which, using the well-known relation $\int_{-\infty}^{\infty}dx \exp(-a x^2 + bx +c) = \sqrt{\pi/a} \,\exp(b^2/(4a) + c)$, can be evaluated as
\begin{align}
	g(i,i') &= \big( \vec{v}_i \cdot \vec{v}_{i'} \big) \sqrt{\frac{2\Delta\omega_i \Delta\omega_{i'}}{(\Delta\omega_i)^2+(\Delta\omega_{i'})^2}} \exp\left[ -\frac{(\omega_{0i}-\omega_{0i'})^2}{4((\Delta\omega_i)^2+(\Delta\omega_{i'})^2)} \right] \\
	&\hspace{0.5cm}\times \exp\left[ -\frac{(\Delta\omega_i)^2 (\Delta\omega_{i'})^2}{(\Delta\omega_i)^2+(\Delta\omega_{i'})^2} (t_{0i}-t_{0i'})^2\right] \exp\left[ -\ii \frac{\omega_{0i}(\Delta\omega_{i'})^2 + \omega_{0i'}(\Delta\omega_i)^2}{(\Delta\omega_i)^2 + (\Delta\omega_{i'})^2} (t_{0i}- t_{0i'}) \right].
\end{align}

\end{document}